\documentclass{article}
\pdfoutput=1
\usepackage{amsmath,amssymb,graphicx,mlspconf}
\usepackage{booktabs}
\usepackage{cite}
\usepackage{microtype}
\usepackage{url}
\usepackage{afterpage}
\def\thebibliography#1{\section{References}\list
 {[\arabic{enumi}]}{\settowidth\labelwidth{[#1]}\leftmargin\labelwidth
 \advance\leftmargin\labelsep \itemsep 1pt plus 0.5pt
 \parsep 0pt
 \usecounter{enumi}}\def\newblock{\hskip .11em plus .33em minus .07em}
 \sloppy\clubpenalty4000\widowpenalty4000\sfcode`\.=1000\relax}

\copyrightnotice{979-8-3195-0884-3/26/\$31.00 {\copyright}2026 IEEE}
\toappear{2026 IEEE International Workshop on Machine Learning for Signal Processing, Sep.\ 28-- Oct.\ 1, 2026, Atlanta, USA}
\title{Subspace Learning with Interval-Censored Likelihoods for Dequantizing Percept\textsuperscript{\texttrademark} PC LFP Snapshots}
\name{\begin{tabular}{@{}c@{}}%
    Shreesh Karjagi$^{\star}$%
    \qquad Elif Ceren Fitoz$^{\star}$%
    \qquad Maryam Khalid$^{\dagger}$%
    \qquad Tanya Nauvel$^{\dagger}$ \\
    Parisa Sarikhani$^{\star}$%
    \qquad Helen S. Mayberg$^{\dagger}$%
    \qquad Christopher J. Rozell$^{\star}$%
    \qquad Sankaraleengam Alagapan$^{\star}$%
\end{tabular}}
\address{%
    $^{\star}$ School of Electrical and Computer Engineering, \\%
    Georgia Institute of Technology, Atlanta, GA 30332 \\%
    $^{\dagger}$ Nash Family Center for Advanced Circuit Therapeutics, \\%
    Icahn School of Medicine at Mount Sinai, New York, NY 10019\\%
}
\begin{document}
\ninept
\raggedbottom

%\raggedbottom
\maketitle

%%
% tldr;  When measurements are coarsely rounded, the right move is to treat each one as an interval. Low-rank estimation under an interval-censored likelihood recovers structure that unconstrained & higher-capacity models destroy.
%%

\begin{abstract}
%%%C1:sufficient background & why we care.
%%%C2:exact numbers.
%%%C3:snthetic ground truth explicit.
%%%C4:ADC specs.
%%%C6/C7:setup sentence. 
%%%C8:downstream impact.
%%%C9:no recommendation.
%R A1 [bKF4-1]: expand FOOOF at first use below 
Implanted neurostimulators that sense local field potentials now enable chronic electrophysiology based biomarker tracking in patients at home. The Medtronic Percept\textsuperscript{\texttrademark} PC, the only commercially available sensing-enabled deep brain stimulation (DBS) device, stores spectral amplitudes as 16-bit integers at approximately $0.1\,\mu$V per bit (quantum $q \approx 0.11\,\mu$Vp). At frequencies where the true amplitude spans only a few quantization levels, consecutive bins round to the same stored value. Standard spectral parameterization (FOOOF, fitting oscillations and one over f), which separates periodic peaks from the aperiodic $1/f$ activity, treats every value as exact and fits oscillatory peaks to these plateaus. Because these spectra feed clinical biomarker pipelines and spectral foundation models for symptom decoding, spurious peaks can corrupt downstream inference. Across 9{,}438 spectra from 14 hemispheres in 7 subcallosal cingulate DBS patients, 20.6\% of peaks detected at [2,\,45]\,Hz have no match in ground truth synthesized by quantizing clean in-clinic BrainSense\textsuperscript{\texttrademark} recordings, while aggregate beta band power and the aperiodic exponent are preserved. We formalize dequantization as interval-censored subspace estimation and compare five classes of correction methods. Quantized probabilistic PCA is the only tested method that reduces the spurious rate (20.6\% to 18.3\%) while preserving true peak detection and keeping noise floor RMSE below $q/\sqrt{12}$. 

\end{abstract}
\section{Introduction}
\label{sec:intro}

%%%C10:start higher level, not a rehash. LFPs for treatment, spectral features, consequence of errors, then quantization.
%%%C11:use technical language "quantization"
%%%C12:cite white paper, use Medtronic terms
%%%C13/C14:FOOOF in context, quantization is the problem not FOOOF
%%%C15:no "run," use "plateau"
%%%C16:artifact worst when amplitude ~ q
%%%C17:delete subspace motivation (moved to Methods, C23)
%%%C18:clarify cohort reference
%%%C19:clear goal, formalization, contribution statement

Sensing-enabled deep brain stimulation (DBS) devices now record local field potentials (LFPs) chronically in patients at home, providing objective electrophysiological measures across movement disorders, epilepsy, and psychiatric conditions~\cite{brontestewart2025,provenza2024,alagapan2023, fitoz2026}. Spectral features extracted from these recordings, including oscillatory peak frequency, band power, and the aperiodic exponent, form the basis of clinical biomarker tracking~\cite{alagapan2023,fitoz2026, veerakumar2019} and spectral foundation models for symptom decoding~\cite{merk2025}. Any systematic distortion in the stored spectrum could propagate directly into these features.

\begin{figure}[!t]
\centering
\includegraphics[width=\columnwidth]{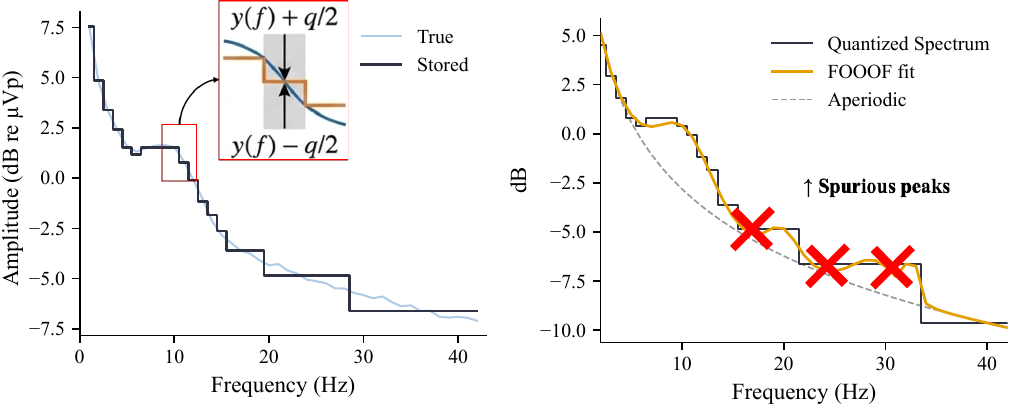}
\caption{Quantization artifact (schematic) and its effect on spectral parameterization. Left: true versus stored spectrum in $\mu$Vp. Right: FOOOF fit to the quantized spectrum in dB. Red crosses mark spurious peaks.}
\label{fig:schematic_visualizationV3}
\end{figure}

The Medtronic Percept\textsuperscript{\texttrademark} PC is the only commercially available sensing-enabled DBS device. Its LFP Snapshot feature stores spectral amplitude values in $\mu$Vp, with each value rounded to the nearest multiple of a device-specific quantum $q \approx 0.11\,\mu$Vp (16-bit storage at approximately $0.1\,\mu$V per bit)~\cite{medtronic_wp}. Quantization is inherent in any digital recording system. At this storage resolution, it collapses consecutive frequency bins to the same stored value wherever the true spectral amplitude varies by less than $q$, producing plateaus of identical amplitudes (Figure~\ref{fig:schematic_visualizationV3}, left). The artifact is most severe at frequencies where the spectral amplitude is comparable to the quantization step, which in subcallosal cingulate recordings occurs above approximately 30\,Hz.

Spectral parameterization, which decomposes a neural power spectrum into periodic oscillatory peaks and an aperiodic $1/f$ activity, has become standard for extracting biomarkers from chronic LFP recordings. FOOOF~\cite{donoghue2020} is the most widely used tool for this decomposition. Like any parameterization that treats spectral values as exact measurements, FOOOF is sensitive to systematic distortions in the input. When quantization produces a plateau of identical bins above the fitted aperiodic curve, the parameterization reports an oscillatory peak that is not present in the underlying signal (Figure~\ref{fig:schematic_visualizationV3}, right). This artifact appears in all seven patients in our dataset (Section~\ref{sec:data}).

Our goal is to recover pre-quantization spectral amplitudes so that downstream parameterization operates on values closer to the true signal. We formalize this as interval-censored subspace estimation and compare five classes of correction methods on 9,438 spectra from seven subcallosal cingulate DBS patients.

\section{Background and Problem Formulation}
\label{sec:problem}

%%%C22:renamed, add background literature context didnt seperate due to space limitations & addressed some issues in into intro
In movement disorders, subthalamic beta power drives adaptive DBS~\cite{brontestewart2025}. In treatment-resistant depression, subcallosal cingulate spectra yield longitudinal recovery biomarkers~\cite{alagapan2023,fitoz2026} and aperiodic slope tracks symptom severity on the timescale of hours~\cite{veerakumar2019}. All of these biomarkers depend on the fidelity of the stored spectrum. Because each stored value is the result of rounding, the true value is not lost entirely but known to lie within a half-quantum interval around it. Interval-censored inference formalizes this structure by treating each observation as constraining the true value to its rounding interval~\cite{buettner2014,lan2014}, and Sanger et al.~\cite{sanger2025} showed that the first principal component of windowed Percept\textsuperscript{\texttrademark} PSDs captures the shared background spectral shape, separating it from transient corruption. This subspace structure is the prior that lets interval-censored methods disambiguate within the rounding interval.

Any analog-to-digital converter rounds continuous values to discrete levels. In the Percept\textsuperscript{\texttrademark} PC, each spectral amplitude is stored as the nearest multiple of $q$. Clean snapshot spectra are not available because the device discards the time-domain data after computing each snapshot~\cite{medtronic_wp}, so we approximate the forward model by applying the same rounding to clean power spectral densities (PSDs) derived from in-clinic BrainSense\textsuperscript{\texttrademark} LFP recordings. Let $s(f) \geq 0$ be the true spectral amplitude at bin $f$ and $q > 0$ the quantization step. The device stores
\begin{equation}
y(f) = \mathrm{round}\!\left(s(f)/q\right) \cdot q,
\label{eq:quantize}
\end{equation}
which constrains the true value to $s(f) \in [\max(0, y(f) - q/2),\; y(f) + q/2)$. For $F$ bins and $N$ spectra the observation matrix $\mathbf{Y} \in \mathbb{R}^{N \times F}$ constrains the true matrix $\mathbf{S}$ elementwise. Spectra from the same recording site share shape and vary in amplitude at a few frequencies (the first five principal components capture $>$99\% of variance in every hemisphere), so we write $\mathbf{S} = \mathbf{1}\boldsymbol{\mu}^\top + \mathbf{Z}\mathbf{W}^\top + \mathbf{E}$ with $\mathbf{Z} \in \mathbb{R}^{N \times K}$ the latent scores, $\mathbf{W} \in \mathbb{R}^{F \times K}$, $K \ll F$, and $\mathbf{E}$ isotropic residual noise distinct from quantization. In our data $K = 5$ suffices ($F = 129$, $N = 366$ to $936$ per hemisphere), and results are stable across $K \in \{3, 5, 7\}$.

\section{Methods}
\label{sec:methods}
%%%C23:rationale for method selection (absorbs C17 motivation)
%%%C25:define standardization
%%%C26:identify CVXPY and OSQP as software
%%%C27:complete Q-PPCA opening sentence
%%%C28:define GPLVM

%R B3 [bKF4-2]: attach a \footnote to "$K = 5$."
%R   \footnote{Iteration counts are convergence ceilings, not tuned: Q-PPCA stops at relative change
%R   $<10^{-4}$ (cap 30), QMF at relative log-likelihood $<10^{-5}$ (cap 200), the DAE by early stopping
%R   (patience 10, cap 100); Adam uses lr $10^{-3}$. Results are stable across $K\in\{3,5,7\}$.}
The interval-censored subspace formulation admits a range of estimators that differ in how they use the quantization intervals and the low-rank prior. We organize our comparison around these two axes. Single-spectrum smoothing (SG, SG-sel) tests whether local polynomial filtering can remove plateaus without cross-spectrum information. Unconstrained subspace projection (SVD) tests whether a learned PCA basis alone suffices. Constrained subspace estimation (SCCD) adds hard interval bounds to the projection. Probabilistic latent variable models with interval-censored likelihoods (Q-PPCA, QMF) test whether a statistical model of the quantization process improves recovery. A denoising autoencoder (DAE) provides a nonlinear baseline that learns a direct mapping from quantized to clean spectra without explicit interval constraints. All subspace methods use $K = 5$.\footnote{Iteration counts are convergence ceilings: Q-PPCA stops at relative parameter change $<10^{-4}$ (cap 30 EM steps), QMF at relative log-likelihood change $<10^{-5}$ (cap 200), and the DAE by early stopping (patience 10, cap 100 epochs). Adam uses learning rate $10^{-3}$.}

\subsection{Savitzky-Golay smoothing (SG, SG-sel)} SG is a second-order polynomial filter~\cite{savitzky1964} with a 5-bin window applied in dB. Bins with zero amplitude are temporarily filled by linear interpolation from neighboring positive bins, the filter runs over the filled spectrum, and the originally zero bins are restored to zero. SG-sel applies the same filter only to plateau bins, where a plateau is a run of two or more consecutive bins whose dB values lie within $10^{-3}$ dB of the first bin.

\subsection{Unconstrained subspace projection (SVD)}
The test spectrum is standardized with training statistics, projected onto the $K$ dimensional PCA basis of the unquantized training fold, and reconstructed. No interval constraints.

\subsection{Subspace constrained convex dequantization (SCCD)}
%R A1 [bKF4-1]: expand MAP"
Maximum a posteriori (MAP) estimator of PCA scores under a Gaussian prior with hard interval constraints. Let $\mathbf{B} \in \mathbb{R}^{K \times F}$ be the PCA basis of standardized training spectra (each frequency bin centered
by its training-set mean $\mu_f$ and scaled by its training-set
standard deviation $\sigma_f$), and $\lambda_k$ the eigenvalues of
the standardized training covariance. The reconstruction is $\hat{s}(f) = \mu_f + \sigma_f \sum_k c_k B_{kf}$ with scores $\mathbf{c}$ solving
\begin{equation}
\min_{\mathbf{c}} \sum_{k=1}^{K} \frac{c_k^2}{\lambda_k}
\quad \text{s.t.} \quad
\ell_f \leq \hat{s}(f) \leq u_f,\; f \in \mathcal{F}_c,
\label{eq:sccd}
\end{equation}
where $\mathcal{F}_c$ is the set of bins in $[1, 42]$\,Hz with positive stored amplitude. Bins outside $\mathcal{F}_c$ are reconstructed from the same $\mathbf{c}$ without constraint enforcement, and all values are clipped at zero. The quadratic program (QP) is specified using the convex modeling library CVXPY~\cite{diamond2016} and solved with the QP solver OSQP~\cite{osqp}. If the QP fails to reach an optimal solution we re-solve with nonnegative slacks on the interval constraints, penalized at $10^3$.

\subsection{Quantized probabilistic PCA (Q-PPCA)}
%R WD2B found Sec. 3.4 hard to follow. The equation itself is correct, so the main fix is to add clearer plain-language guidance around it.
%R A3 [WD2B]: This paragraph currently says "The moments in Eq.~\eqref{eq} are exact..." before Eq.~(3) has been introduced. Remove that clause here. Then, after the equation, add the clarification that these moments are exact for univariate truncated normals.
%R B4 [WD2B]: Make the objects more explicit. After "The exact row posterior is a multivariate truncated normal," add that this is the distribution over the true amplitudes given their quantization intervals. Also gloss the PPCA predictive mean, $\mathbf{W}\mathbf{M}^{-1}\mathbf{W}^\top(\mathbf{y}_i-\boldsymbol{\mu})+\boldsymbol{\mu}$, as the value PPCA would predict if the observation were exact.
Q-PPCA extends probabilistic PCA~\cite{tipping1999} to handle interval-censored observations arising from quantization. Buettner et al.~\cite{buettner2014} handle left censored gene expression data with a Gaussian process latent variable model and a probit approximation to the censored likelihood. We use a simpler model with exact truncated normal moments instead of assumed density filtering, and our censoring is two-sided (interval) rather than one-sided (left). The generative model is $\mathbf{z}_i \sim \mathcal{N}(\mathbf{0}, \mathbf{I}_K)$, $\mathbf{s}_i = \mathbf{W}\mathbf{z}_i + \boldsymbol{\mu} + \boldsymbol{\varepsilon}_i$ with $\boldsymbol{\varepsilon}_i \sim \mathcal{N}(\mathbf{0}, \sigma^2\mathbf{I})$, observed through the quantizer in Eq.~\eqref{eq:quantize}. Training alternates between an approximate E-step and a standard M-step. The exact row posterior, the distribution over the true amplitudes given their quantization intervals, is a multivariate truncated normal. We approximate it by a product of per-bin truncated normals centered at the PPCA predictive mean, with per-bin predictive variance $\sigma^2\,\mathrm{diag}(\mathbf{W}\mathbf{M}^{-1}\mathbf{W}^\top) + \sigma^2$ where $\mathbf{M} = \mathbf{W}^\top\mathbf{W} + \sigma^2\mathbf{I}$. The factorization across bins and the use of the uncensored observation in the conditioning are the two approximations.
\begin{equation}
\mathbb{E}[s_{ij}] = \hat{m}_{ij} + \sqrt{v_j}\,\frac{\phi(\alpha_{ij}) - \phi(\beta_{ij})}{\Phi(\beta_{ij}) - \Phi(\alpha_{ij})},
\label{eq:trunc}
\end{equation}

%added init rationale (WD2B)
where $\hat{m}_{ij}$ is the PPCA predictive mean (the value predicted if the observation were exact), $v_j$ the per-bin predictive variance, and $\alpha_{ij}, \beta_{ij}$ the standardized interval bounds. Here $\phi$ and $\Phi$ are the standard-normal pdf and cdf, and the moments in Eq.~\eqref{eq:trunc} are exact for univariate truncated normals. Up to 30 EM iterations with tolerance $10^{-4}$, initialized from the SVD of the centered quantized training matrix, which only warm-starts the EM and is then re-estimated on the quantized data, unlike the fixed unquantized basis that SVD and SCCD use.

At correction time the parameters are fixed. The corrected bin value is the per-bin truncated normal mean under the predictive distribution conditioned on the test bin's quantization interval. Exact inference under test time censoring would require iterating.

\subsection{Quantized matrix factorization (QMF)}
We adapt the quantized matrix completion formulation of Lan et al.~\cite{lan2014} to the Percept\textsuperscript{\texttrademark} PC setting. The bin boundaries are known from the device and all frequency bins are observed, so we drop their boundary estimation and missing data machinery. We enforce rank by factorization rather than nuclear norm relaxation. We factorize $\mathbf{S} \approx \mathbf{U}\mathbf{V}^\top + \mathbf{1}\boldsymbol{\mu}^\top$ by maximizing
\begin{equation}
\sum_{ij} \log\bigl[\Phi(\beta_{ij}) - \Phi(\alpha_{ij})\bigr],
\label{eq:qmf}
\end{equation}
where $\alpha_{ij}, \beta_{ij}$ are the standardized bin boundaries under the low rank model with noise scale $\sigma = q/2$ fixed throughout optimization. We optimize with Adam~\cite{kingma2015} (learning rate $10^{-3}$, elementwise gradient clipping to $[-5, 5]$) for up to 200 iterations or until relative log likelihood change falls below $10^{-5}$. At test time $\mathbf{V}$ and $\boldsymbol{\mu}$ are fixed, a new row $\mathbf{u}$ is obtained by least squares projection onto the column space of $\mathbf{V}$, and the reconstruction is clipped bin by bin to the quantization interval.

\subsection{Denoising autoencoder (DAE)}
A feed-forward autoencoder~\cite{hinton2006} ($F \to 256 \to 256 \to F$, ReLU, dropout 0.1) trained in the denoising configuration of Vincent et al.~\cite{vincent2008} where quantization serves as deterministic corruption and the clean spectrum is the target. Inputs and targets are standardized per-bin using clean training statistics. Adam~\cite{kingma2015} with learning rate $10^{-3}$, weight decay $10^{-4}$, batch 64, up to 100 epochs with early stopping (patience 10). At test time the output is destandardized and clipped at zero. No interval constraint is applied.

\section{Experimental Setup}
\label{sec:experiments}
%%%C29:renamed to Experimental Setup
%%%C30:synthetic protocol explicit

\subsection{Data}
\label{sec:data}
Seven patients (S1 to S7) with bilateral subcallosal cingulate DBS for treatment-resistant depression were included in this analysis, giving 14 hemispheres. All participants provided informed consent under an IRB-approved protocol. The LFP Snapshot feature stores only the averaged amplitude spectrum and discards the underlying time-domain data~\cite{medtronic_wp}, so the original time series from which each snapshot spectrum is derived is not available. To construct ground truth, we use BrainSense\textsuperscript{\texttrademark} recordings collected in-clinic at 250\,Hz, with 45 to 114 recordings per hemisphere across clinic visits. Each recording is cut into 30\,s segments, yielding 366 to 936 per hemisphere and 9{,}438 in total. Clean amplitude spectra are computed via Welch's method (256-sample Hann window, 50\% overlap), producing 129 bins from 0 to 125\,Hz. Amplitudes are in $\mu$V peak ($\mu$Vp): the per-bin Fourier amplitude $\sqrt{2\,S_{xx}(f)\,\Delta f}$ from the one-sided density $S_{xx}$, not a power spectral density. We then apply the device quantization model (Eq.~\ref{eq:quantize}) with $q = 0.11\,\mu$Vp to each clean spectrum. Each clean spectrum serves as ground truth, and its quantized counterpart serves as the test input for all correction methods. All evaluation in this paper uses this synthetic protocol. Code is available at \textit{\mbox{\url{https://github.com/shreeshkarjagi/ic-subspace-learning}}}.

\subsection{Evaluation protocol}
Raw is the uncorrected quantized reference. SG and SG-sel process each spectrum independently and use no training data. Subspace methods (SVD, SCCD, Q-PPCA, QMF) and DAE are evaluated under two cross-validation schemes. LOO holds out all segments from one recording, learns the model from the remaining recordings of the same hemisphere, and applies correction to each held-out segment. LOPO holds out all hemispheres from one patient, learns the model from the other six patients, and applies correction to all held-out segments. SVD and SCCD learn their basis from unquantized training spectra. Q-PPCA and QMF train on the quantized training matrix. The DAE trains on quantized inputs with clean targets.

\subsection{Spectral parameterization and metrics}
We run FOOOF~\cite{donoghue2020} on clean and corrected spectra at [2,\,45]\,Hz with peak width limits [2,\,8]\,Hz, maximum 4 peaks, minimum height 0.1\,dB, and fixed aperiodic mode. Peaks are matched between clean and corrected fits by center frequency within 2\,Hz using greedy nearest neighbor. We report RMSE in $\mu$Vp across three frequency regimes (well-resolved WR 1 to 8\,Hz, correctable CR 8 to 30\,Hz, noise floor NF 30 to 42\,Hz). Uniform quantization predicts $q/\sqrt{12} \approx 0.032\,\mu$Vp. Secondary metrics are spectral SSIM (7-bin sliding window in dB), quantization consistency (fraction of corrected bins that re-quantize to the stored value), and median plateau count. We also report peak detection rate and spurious rate. Significance of the spurious-rate reduction is assessed by a paired Wilcoxon signed-rank test across the 14 hemispheres and a 95\% bootstrap confidence interval resampling the 9{,}438 spectra ($5{\times}10^3$ draws).

\section{Results}
\label{sec:results}

\subsection{Quantization contaminates peak detection}

\begin{figure*}[!t]
\centering
\includegraphics[width=\textwidth]{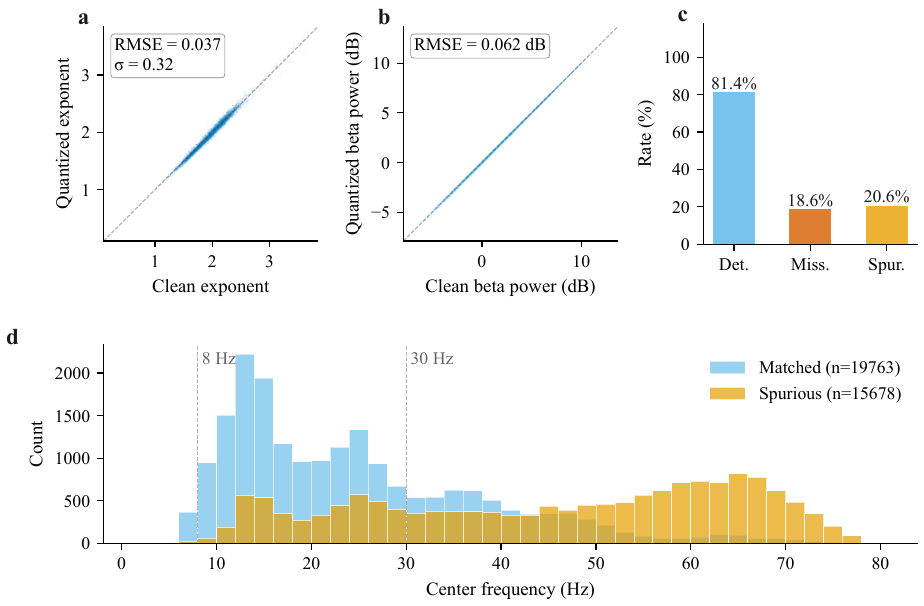}
\caption{FOOOF on quantized versus clean spectra: exponent (a), beta band power (b), and peak rates (c) at [2,\,45]\,Hz. Spurious and matched center frequencies (d) shown at [2,\,80]\,Hz for full spectrum context.}
\label{fig:specparam_impact}
\end{figure*}

The aperiodic exponent is robust to quantization. On raw spectra the exponent RMSE at [2,\,45]\,Hz is 0.037 against a natural cohort standard deviation of 0.32. Band power RMSEs in dB are below 0.07 for theta through beta but reach 0.176 in low gamma. The FOOOF fit $R^2$ drops only from 0.991 to 0.987 (Figure~\ref{fig:specparam_impact}a,b). A pipeline that tracks only aggregate parameters would see nothing wrong.

But 20.6\% of peaks reported on raw quantized spectra have no match in the clean fit within 2\,Hz (Figure~\ref{fig:specparam_impact}c). Median spurious power is 0.23\,dB above the fitted slope, 90th percentile 0.46\,dB. These detections concentrate above 30\,Hz where amplitude spans only a few quantization levels (Figure~\ref{fig:specparam_impact}d). The spurious rate depends sharply on the FOOOF fit range: 15.5\% at [2,\,30]\,Hz, 20.6\% at [2,\,45]\,Hz, and 44.2\% at [2,\,80]\,Hz, while exponent RMSE stays below 0.06 through [2,\,60]\,Hz but breaks 0.10 at [2,\,80]\,Hz.

\subsection{Spurious peak reduction after correction}

\begin{figure*}[!t]
\centering
\includegraphics[width=\textwidth]{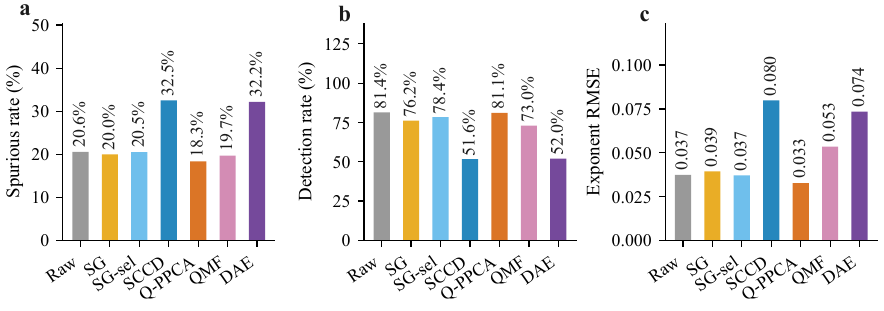}
\caption{Spurious rate (a), detection rate (b),  and exponent RMSE (c) after correction at [2,\,45]\,Hz. SVD omitted for scale.}
\label{fig:spurious_reduction}
\end{figure*}

We applied each method to all 9{,}438 spectra and ran FOOOF on the corrected output (Figure~\ref{fig:spurious_reduction}). SG and SG-sel shed true and spurious peaks at similar rates, so the spurious rate stays near 20\%. QMF trades 8 points of detection for less than a point on the spurious rate. SCCD and DAE both make things worse (spurious rates 32.5\% and 32.2\%, detection near 52\%). SVD destroys the aperiodic fit (exponent RMSE $>$1).

Q-PPCA is the only method that removes more artifacts than it introduces. The spurious rate drops 2.24 points, from 20.6\% to 18.3\%, in all 14 hemispheres (paired Wilcoxon $p = 1.2\times10^{-4}$, 95\% bootstrap CI $[1.9, 2.6]$ over 9{,}438 spectra), while detection holds at 81.1\% from 81.4\%, exponent RMSE drops from 0.037 to 0.033, and exponent bias moves from $+0.009$ to $+0.004$. Remaining spurious peaks sit above 30\,Hz where each bin spans one to three quantization levels and no subspace method can resolve whether a given bin should read 0.11 or 0.22\,$\mu$Vp.

\begin{figure*}[!t]
\centering
\includegraphics[width=\textwidth]{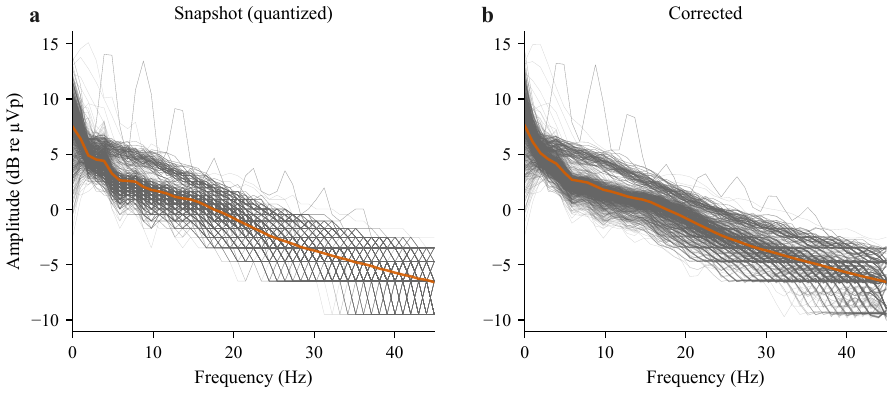}
\caption{All snapshots from one hemisphere. Stored (a), Q-PPCA corrected (b).}
\label{fig:exemplar}
\end{figure*}

\subsection{Spectrum recovery and consistency}
\begin{table}[!t]
\caption{RMSE ($\mu$Vp) by regime across 9{,}438 spectra. SG and SG-sel use no training data, so their LOO and LOPO columns are identical by construction.}
\label{tab:rmse}
\centering
\small
\setlength{\tabcolsep}{4pt}
\begin{tabular}{@{}lccc c ccc@{}}
\toprule
& \multicolumn{3}{c}{LOO} & & \multicolumn{3}{c}{LOPO} \\
\cmidrule{2-4} \cmidrule{6-8}
Method & WR & CR & NF & & WR & CR & NF \\
\midrule
Raw    & \textbf{0.031} & 0.032 & 0.031 & & \textbf{0.031} & 0.032 & 0.031 \\
SG     & 0.125 & 0.052 & 0.028 & & 0.125 & 0.052 & 0.028 \\
SG-sel & 0.045 & 0.037 & 0.029 & & 0.045 & 0.037 & 0.029 \\
SVD    & 1.515 & 0.449 & 0.117 & & 0.738 & 0.294 & 0.078 \\
SCCD   & 0.243 & 0.123 & 0.044 & & 0.309 & 0.163 & 0.047 \\
Q-PPCA & 0.034 & \textbf{0.032} & \textbf{0.028} & & 0.033 & \textbf{0.032} & \textbf{0.029} \\
QMF    & 0.055 & 0.049 & 0.033 & & 0.056 & 0.051 & 0.034 \\
DAE    & 0.345 & 0.133 & 0.040 & & 0.428 & 0.147 & 0.048 \\
\bottomrule
\end{tabular}
\end{table}

Table~\ref{tab:rmse} reports RMSE by regime. Raw RMSE is near 0.032\,$\mu$Vp in all three regimes, the uniform quantization floor $q/\sqrt{12}$. Between 1 and 8\,Hz the signal is large enough that per-bin relative error is already tiny, and every method except Q-PPCA degrades Raw. Above 30\,Hz each bin spans one to three quantization levels. Q-PPCA reaches 0.028\,$\mu$Vp LOO and 0.029\,$\mu$Vp LOPO at the noise floor, below $q/\sqrt{12}$. SG reaches the same 0.028 but by polynomial smoothing that destroys the well-resolved band (0.125\,$\mu$Vp at WR). Q-PPCA is the only method that improves the noise floor while keeping WR within 0.003\,$\mu$Vp of Raw. LOO and LOPO differ by at most 0.001\,$\mu$Vp for Q-PPCA in any regime, so the learned subspace transfers to unseen patients.

SVD and SCCD move in opposite directions under LOPO. SVD improves because a six-patient basis is less overfit than a single hemisphere one. SCCD degrades because its box constraints become infeasible on more bins when the test spectrum sits outside the training covariance. Apart from SVD, DAE is worst at WR (0.345\,$\mu$Vp LOO, 0.428\,$\mu$Vp LOPO), an order of magnitude above Raw.

\begin{table}[!t]
\caption{Secondary metrics under LOO.}
\label{tab:secondary}
\centering
\small
\begin{tabular}{@{}lccc@{}}
\toprule
Method & SSIM & Consist. & Plateaus \\
\midrule
Raw    & 0.769          & \textbf{100\%} & 14 \\
SG     & 0.787          & 89\%           & 5  \\
SG-sel & 0.779          & 98\%           & 5  \\
SVD    & 0.720          & 60\%           & 1  \\
SCCD   & 0.895          & 71\%           & 0  \\
Q-PPCA & 0.837          & $\geq$99.9\%   & 2  \\
QMF    & 0.831          & 87\%           & 8  \\
DAE    & \textbf{0.920} & 77\%           & 0  \\
\bottomrule
\end{tabular}
\end{table}

Of the subspace methods only Q-PPCA keeps consistency above 99\% and still improves SSIM over Raw (Table~\ref{tab:secondary}). SCCD has the highest SSIM among subspace methods (0.895) but stays within the quantization interval on only 71\% of bins. A method that moves values outside their stored bin that often is smoothing more than dequantizing. DAE scores the highest SSIM overall (0.920) because its smooth output in dB is forgiving to the 7 bin SSIM window, but its consistency is only 77\% and its RMSE is an order of magnitude above Raw. SSIM alone is misleading here. Q-PPCA drops the median plateau count from 14 to 2, removing the staircase without flattening the spectrum. Figure~\ref{fig:exemplar} shows all snapshots from one hemisphere before and after Q-PPCA correction and SG smoothing. On one hemisphere (S6 left, 936 segments), Q-PPCA RMSE varies by at most 0.003\,$\mu$Vp across $K \in \{3, 5, 7\}$ in all three regimes, while SVD WR varies by 1.8\,$\mu$Vp, confirming that Q-PPCA is stable across rank.

%%%C32:why beta is a salient special case

\section{Discussion}
\label{sec:discussion}

Quantization at the Percept\textsuperscript{\texttrademark}'s ${\sim}0.11\,\mu$Vp resolution produces an artifact that standard parameterization cannot distinguish from a real oscillation at frequencies where the true spectral amplitude spans only a few quantization levels. The aperiodic exponent and beta band power are preserved, so aggregate metrics do not flag the contamination. The problem shows up in single-peak detection when the fit range extends into frequencies where each bin spans a few quantization levels.\looseness=-1

The right correction depends on the task. Beta band (13--30\,Hz) oscillatory peaks are the primary spectral biomarker tracked by the Percept\textsuperscript{\texttrademark} PC in movement disorders and treatment-resistant depression, making beta peak detection the most common clinical use of these spectra. For studies whose only use of the spectrum is beta peak detection, restricting FOOOF to [2,\,30]\,Hz is sufficient and requires no correction at all. For pipelines that treat the full spectrum as a feature vector, including spectral foundation models for symptom decoding~\cite{merk2025}, restricting the fit range discards the frequencies that quantization corrupts most. In that setting Q-PPCA is the only tested method that reduces the spurious rate while keeping detection, noise floor RMSE, and bin consistency intact.\looseness=-1

Interval constraints alone do not explain Q-PPCA's advantage. SVD ignores the quantization intervals and overshoots. SCCD enforces hard box constraints but has no likelihood to mediate conflicts between the $K{=}5$ subspace and the 129 interval bounds, so when the test spectrum sits outside the training covariance the QP goes infeasible and the slack fallback lets the reconstruction leave its stored bin. QMF has an interval censored likelihood but no row prior, so its factors overfit and it trades detection for a marginal spurious reduction. DAE fails differently: MSE training produces a smooth conditional mean that distorts the spectral shape, and FOOOF fits peaks to the distortion. Q-PPCA combines a censored likelihood and a learned row prior, and among the methods tested that combination yields the best trade-off between spurious reduction and detection preservation.\looseness=-1

The contribution is the empirical finding that quantization produces spurious peaks that could affect downstream spectral pipelines, and a systematic comparison showing that only one class of correction helps. Q-PPCA extends PPCA~\cite{tipping1999} with truncated normal E-step moments, related to the censored PCA of Buettner et al.~\cite{buettner2014}. QMF adapts the quantized matrix completion model of Lan et al.~\cite{lan2014}. The interval censored subspace framing connects this device-specific problem to a general estimation setting applicable to any brain-computer interface or embedded sensor that stores spectra as quantized amplitudes.\looseness=-1

\section{Conclusion}
\label{sec:conclusion}
%%% C34:fix tone make less negative

Quantization is inherent in any digital spectral representation. At the storage resolution of the Percept\textsuperscript{\texttrademark} PC ($q \approx 0.11\,\mu$Vp), it produces spurious spectral peaks that are invisible to aggregate band power and exponent tracking. We formalized correction as interval censored subspace estimation and compared five methods on 9{,}438 spectra from seven patients. Q-PPCA is the only method that reduces the spurious rate without degrading detection or leaving the stored bin structure.\looseness=-1

Two limitations bound these results. The ground truth is synthesized by quantizing clean streaming PSDs, not by comparing against an independent unquantized snapshot recording, which the device does not store. And the improvement is modest. Most remaining spurious peaks sit above 30\,Hz where the median amplitude is below $1.65\,\mu$Vp but the spectral slope is shallow enough that adjacent bins differ by less than $q$ and round to the same stored value, so no subspace method can resolve them. Quantization aware spectral parameterization, where the aperiodic fit accounts for interval censoring directly, may prove more effective than correcting the spectrum first and fitting second.\looseness=-1

\section{Acknowledgments}
\label{sec:acknowledgments}

The authors thank Scott Stanslaski and Ben Isaacson (Medtronic, Inc.) for technical support. Percept\textsuperscript{\texttrademark} PC devices were provided by Medtronic, Inc. This work was supported by NIH grants UH3NS103550 and UH3NS141080, the Wellcome Leap MCPsych program, and the Hope for Depression Research Foundation. H.S.M. receives consulting and intellectual property licensing fees from Abbott Neuromodulation.\looseness=-1

\raggedbottom
{\small
\bibliographystyle{IEEEbib}
\bibliography{refs}
}
\end{document}